\documentstyle[aps,pra,preprint]{revtex}
\begin{document}
\draft
\title{Calculation of minor hysteresis loops under metastable to stable transformations in vortex matter}
\author{P. Chaddah and M. Manekar}
\address{Low Temperature Physics Laboratory,\\
Centre for Advanced Technology,\\ Indore 452013, India}
\date{\today }
\maketitle
\begin{abstract}
We present a model in which metastable supercooled phase and stable
equilibrium phase of vortex matter coexist in different regions of a sample.
Minor hysteresis loops are calculated with the simple assumption of the two
phases of vortex matter having field-independent critical current densities.
We use our earlier published ideas that the free energy barrier separating 
the metastable and stable phases reduces as the magnetic induction moves 
farther from the first order phase transition line, and that metastable 
to stable transformations occur in local regions of the sample when the 
local energy dissipation exceeds a critical value. Previously reported 
anomalous features in minor hysteresis loops are reproduced, and calculated
field profiles are presented.

\end{abstract}

\section{Introduction}
Supercooled or metastable states have been
reported across first order phase transitions in vortex 
matter \cite{1,2,3,4,5,6,7,8,9,10}. It has also been established that 
isothermal field excursions cause the metastable phase to be converted to 
the stable phase  \cite{5,7,8,10}. We have proposed \cite{11} that the 
isothermal field variations  
provide a fluctuation energy that causes the metastable
supercooled phase to cross the free-energy barrier and transform
to the stable equilibrium phase. 

The experimental techniques used to study the magnetic signatures of such
 transformations in vortex matter are (i) bulk dc measurements using a SQUID
 or a vibrating sample magnetometer which yields the magnetisation M of the
entire sample \cite{5,6,7,8,9,10}; (ii) bulk ac measurements of 
susceptibility which probe a region near the surface of the sample \cite{3};
 and (iii) local measurements of magnetic induction using magneto-optic or 
microhall probes that allow mapping the spatial profile B(x) \cite{1,2,4}. 
The last of these three techniques has been used recently to show that 
different (metastable and equilibrium) phases exist simultaneously in 
different regions of  the sample \cite{1,2}. Similar inference has also been
drawn by studies using the dc magnetisation \cite{5,6} and ac susceptibility 
\cite{3} techniques. Experiments have thus shown that metastable to stable 
transformations occur over local regions,and prompted by these developments 
we have given a formalism to calculate spatially-resolved energy dissipation 
under an isothermal field variation \cite{12}. In this paper we shall use 
this formalism to calculate how the metastable to stable transformation 
progresses inwards from the surface of the sample, under experimentally 
relevant field excursions. We shall also calculate the sample magnetisation 
M and the spatial field profile B(x) as the experimental observables.

\section{Modelling the Peak Effect}

The response of a hard superconductor to external magnetic fields is 
understood in terms of Bean's critical state model (CSM) \cite{13}. 
Bean had assumed that the critical current density J$_C$ is independent of 
field. While detailed agreement with experiments has required introduction 
of various functional forms of J$_C$(B) (see e.g. ref. [14]), much of the
essential physics is captured even by assuming a field-independent J$_C$.

Experiments have recently been addressing the region below and near the 
onset of a peak in J$_C$(B) at B=B$_1$, where a first order phase transition 
is seen in some superconductors\cite{1,2,3,4,5,6,7,8,9,10}. The occurrence 
of this "peak-effect" has been known in various superconductors for a very
long time, but attempts to have a CSM describing this J$_C$(B) have been 
made only recently \cite{15,16}. While our detailed analytical 
model \cite{15} could be used for the subsequent calculations, our focus 
here is to understand whether qualitatively new and anomalous signatures 
in recent experiments \cite{4,5,6,7,8,9,10} can be arising from metastable 
to stable transformations in vortex matter. In this paper we shall use a 
simple Bean-like assumption for the two phases of vortex matter viz.
\begin{eqnarray}
J_C(B)&=& J_1 \mbox{    for  }  B \leq B_1 \nonumber \\
      &=& J_2 \mbox{    for  }  B \geq B_1   
\end{eqnarray}
We stress that, at a fixed temperature, our model has only two constant
 parameters viz. (J$_2$/J$_1$) and B$_1$. We have assumed above that phase 
1 (characterised by J$_1$) is the stable phase for B(x)$\leq B_1$, and phase
2 (characterised by J$_2$) is the stable phase for B(x)$\geq B_1$. 
We recognise that J$_C$ is not a thermodynamic quantity, but is a physical 
property that changes discontinuously across the phase transition.
 Phase 2 can exist for B(x) $\leq B_1$ as a supercooled metastable phase, 
and we shall address this possibility in the next section. In this section 
we shall assume, however, that the free energy barrier surrounding the 
metastable phase drops very sharply as B(x)=B$_1$ is crossed, 
and supercooling or superheating does not occur.

To obtain magnetisation-vs-field (or M-H) curves, we 
consider the sample to be in the form of an infinite slab in parallel field,
as this geometry has the simplest algebra amongst the zero
demagnetization factor cases of infinite cylinders in parallel
field.  We shall also continue with Bean's simplifying assumption of
H$_{C1}$=0 followed usually in the CSM \cite{13,14,15}.

We follow standard procedures \cite{14,15} to solve the CSM, and show in 
figure 1 the envelope M-H curves obtained with equation (1), with the 
parameters B$_1$=1500 mTesla, J$_1$R=4mTesla, and J$_2$R=10 mTesla. 
Here the slab has surfaces at x=$\pm$R, is infinite along the y and z 
directions, and the magnetic field is applied along the z-axis. 
(We shall consider only positive values of x in this paper; there is a 
symmetry about x=0.) Note that the first order transition shows different 
widths in M vs H when measured along the field-increasing and along the 
field-decreasing directions. This is because the shielding current density 
at x is dictated \cite{13,14,15,16} by the local magnetic induction B(x) 
through equation (1), and B(x) is different from the applied field as well 
as different in the field-increasing and field-decreasing cases. 
In figure 2 we plot B(x) for some values of applied field H corresponding 
to the field-increasing and field-decreasing cases. We note that phase 1 
and 2 exist simultaneously in two different regions of the sample. We 
emphasize that there is no metastability because the stable phase 1 exists 
wherever B(x)$\leq B_1$ and the stable phase 2 exists wherever 
B(x)$\geq B_1$. 

The calculation above is for some fixed temperature T$_1$, and we note that 
the phase transition field B$_1$ falls as the temperature T$_1$ rises \cite{17}.

\section{Supercooling and metastable-to-stable transformations}

We now consider that we have applied a field H$_1$ which is smaller than 
B$_1$(T$_1$). But we apply this field at a much higher temperature T$_2$ 
such that H$_1$ is much larger than B$_1$(T$_2$). So, B(x) throughout the 
sample is  larger than B$_1$ at that temperature, and the entire sample is 
in phase 2. We assume further that B(x) is constant at T$_2$. This happens 
if the critical current density J$_2$ in phase 2 vanishes at T$_2$. One can,
 however, also achieve a constant B(x) by applying an external 
field H$_1$+hcos(wt), with (H$_1$-B$_1$(T$_2$)) $>$ h $>$ J$_2$(T$_2$)R, 
and then slowly reducing the amplitude h to zero \cite{14,18}.

We now lower the sample temperature (i.e. field-cool) to T$_1$ such that 
the sample is supercooled and is metastable in phase 2. As discussed in 
references [11,12,17], there is a free energy barrier f$_B$(T) that keeps 
phase 2 metastable, where f$_B$(T) is determined uniquely by B(x) and T. 
The vortex matter in the neighbourhood of x will transform to phase 1 when 
the fluctuation energy P$_d$(x) created by an isothermal field variation 
is larger than [f$_B$(T) - kT].

The field profile B(x) in the field-cooled sample is constant at H$_1$ 
(see figure 3), and we now start lowering the applied field H with the 
temperature fixed at T$_1$. Since the sample is in the supercooled phase 2, 
the shielding currents set up initially will have a magnitude J$_2$. 
The variations in field will cause a fluctuation energy P$_d$(x) given 
by equations (4) and (5) of reference [12], and the vortex matter in the 
neighbourhood of x=x$_0$ will transform to the stable phase at 
P$_d$(x$_0$)= f$_B$(T) - kT =P$_0$. This transformation is triggered 
from the surface \cite{12} and the shielding current magnitude will drop 
to J$_1$ for x $>$ x$_0$. The point x$_0$ moves from (x/R)=1 to (x/R)=0 as
 the applied field is lowered, and B(x) are shown in figure 3 for
 representative values of the applied field. We have used H$_1$=1480 mTesla, 
and P$_0$=18 (mTesla)$^2$. The large (small) slopes of B(x) correspond to
 large (small) magnitudes of the shielding current density, and thus to
vortex matter being in phase 2 (phase 1). From these B(x) one can readily
 calculate \cite{13,14,15}  the sample magnetisation as the field H is
 lowered. In figure 4 we show the minor hysteresis loop (MHL) obtained as 
the applied field is lowered after field-cooling. We show also the
 field-decreasing envelope curve from figure 1. Note that the MHL first 
shoots out above the envelope curve,and then slowly merges from above. 
This nature is in qualitative agreement with  published 
data \cite{7,9,10,19,20}. If we had used the detailed J$_C$(B) of 
reference [15] to model the peak-effect, instead of the simple model 
of equation (1), the peak of the MHL would be less sharp and the merger
 with the envelope curve would be slower. The simple model used to obtain 
figure 4 brings out the qualitative behaviour observed and captures the
 essential underlying origin of anomalous MHLs as being due to phase 
2 being supercooled and the transformation from the metastable phase 2 
to the stable phase 1 occurring progressively deeper into the sample.

We have assumed that P$_0$ (and thus f$_B$(T$_1$)) is constant at 18 
(mTesla)$^2$. f$_B$(T$_1$) is actually dictated by B(x$_0$), and falls 
monotonically as B(x$_0$) moves farther from the phase transition 
line B$_1$(T). As is seen in figure 3, B(x$_0$) varies only by less
 than a few mTesla as the MHL merges with the envelope curve. 
The assumption of a constant P$_0$ over an MHL is thus justified. 
If, however, we field-cool to the same temperature T$_1$ at a lower
 field H$_2$, then f$_B$(T) will be lower \cite{17}. 
This implies that P$_0$(H$_{FC}$=H$_2$) will be smaller than 
P$_0$(H$_{FC}$=H$_1$). We show, in figure 5(a), the MHL for the case when
 the sample was field-cooled to H$_2$=1400 mTesla where P$_0$ is taken 
to be 4.5 (mTesla)$^2$. In figure 5(b) we have taken
H$_{FC}$=H$_3$=1300 mTesla, where f$_B$ must be still lower and is taken 
as P$_0$=2 (mTesla)$^2$. The MHLs again shoot out of the envelope curve,
 but to peak values progressively smaller than in figure 4. The merger of
 the MHLs with the envelope curve also occurs over a progressively narrower 
range of field reduction than in figure 4. This qualitative change in the
 nature of the MHLs with reduction of H$_{FC}$ is also consistent with 
published data \cite{7,9,10,19,20}.    

\section{Conclusion}

We have used the ideas developed in references [11,12,17] to calculate 
the isothermal field-cooled MHLs, and the spatial field profiles B(x).
The model calculation was done, in the spirit of Bean's original 
work \cite{13}, with field-independent critical current densities. 
The only parameters were (J$_2$/J$_1$), and the onset field B$_1$ at 
which the peak effect starts in the field increasing case. We used the 
fact \cite{17} that f$_B$ becomes smaller as B(x) falls below B$_1$,
 and that metastable to stable transformations occur in local regions 
of the sample \cite{12}.

The formalism of reference [12] can similarly be used to calculate MHLs 
after different thermomagnetic histories. We assert here that the simple 
model of equation (1) reproduces qualitative features of various 
observations \cite{5,6,7,8,9,10,19,20} of anomalous MHLs. As was stated 
in the Introduction, more detailed tests of the extent of phase coexistence 
are possible and calculated B(x) can be compared with field profiles
 measured with local probes. Our model also predicts the spatial region 
over which the two phases coexist, and the evolution of these regions 
under isothermal field variation. 

We thank Drs. S.B. Roy and Sujeet Chaudhary for many helpful discussions.

\begin{figure}
\caption{ Field-increasing and field-decreasing envelope M-H curves are
shown, following the model of equation (1), with B$_1$=1500 mTesla.
The crosses indicate applied field values at which B(x) profiles are shown 
in figure 2.}
\end{figure}

\begin{figure}
\caption{ B(x) profiles are shown for (a) field-increasing case at applied 
fields of 1495, 1505, and 1515 mTesla; and (b) field-decreasing case at
 applied fields of 1501, 1498, and 1495 mTesla. The arrows indicate the x 
at which B(x)=1500 mTesla.}
\end{figure}

\begin{figure}
\caption{ The sample is field-cooled in 1480 mTesla, when B(x) is constant 
as shown by the dashed line, and vortex matter is in the metastable phase 2.
 B(x) are shown as the applied field is lowered isothermally to 1475, 1473,
and 1471 mTesla. In the last two fields the vortex matter has transformed 
to the stable phase 1 at x$>$x$_0$, where x$_0$ is indicated by an arrow.}
\end{figure}

\begin{figure}
\caption{ MHL obtained after field-cooling at 1480 mTesla is shown by the 
solid line. It overshoots the envelope curve, and merges with it slowly 
from above. The circle indicates the starting point of the MHL, 
with M=0 corresponding to the constant B(x).}
\end{figure}

\begin{figure}
\caption{ Same as figure 4 except that field-cooling was done at (a) 
1400 mTesla; and (b) 1300 mTesla.}
\end{figure}

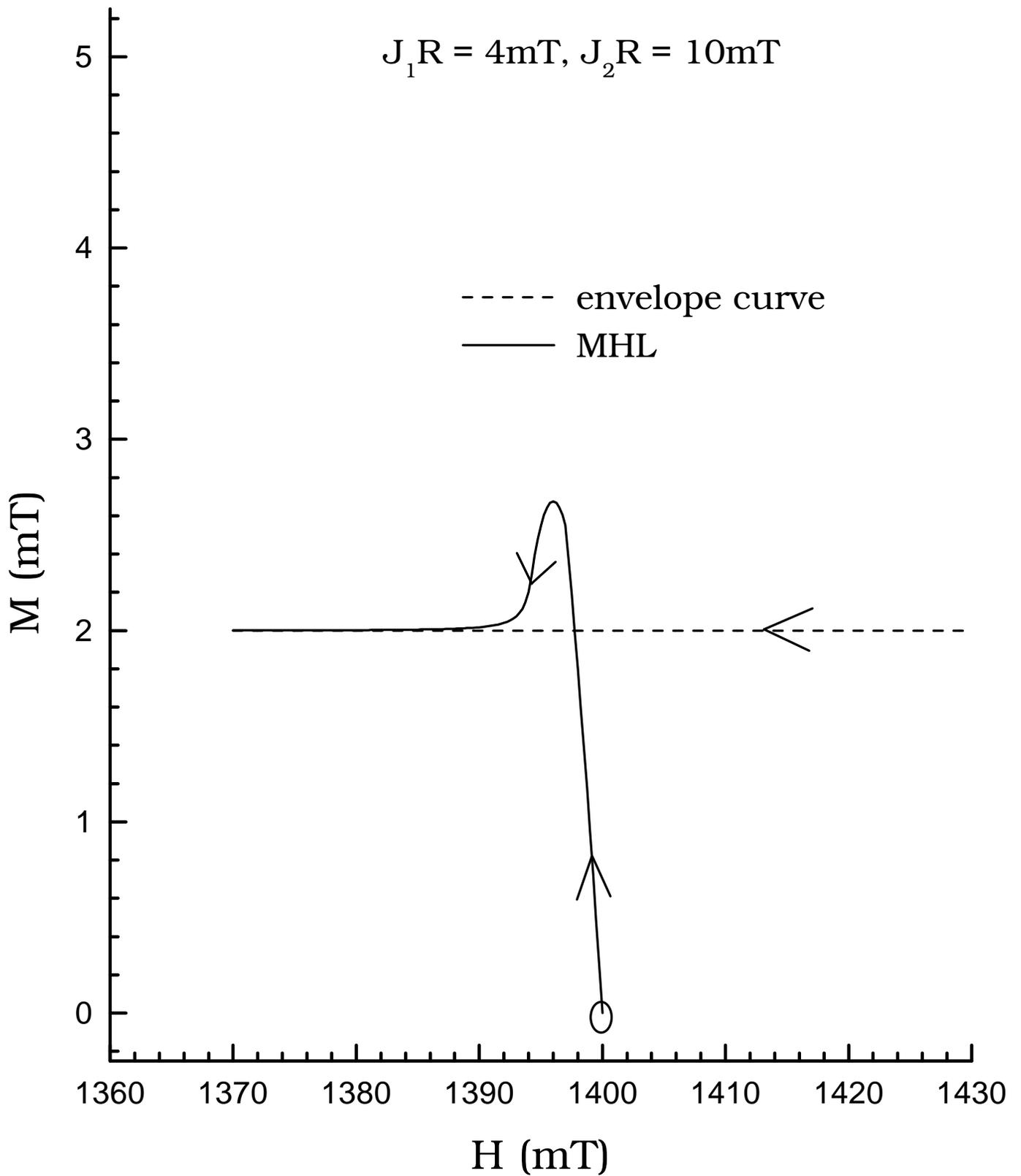
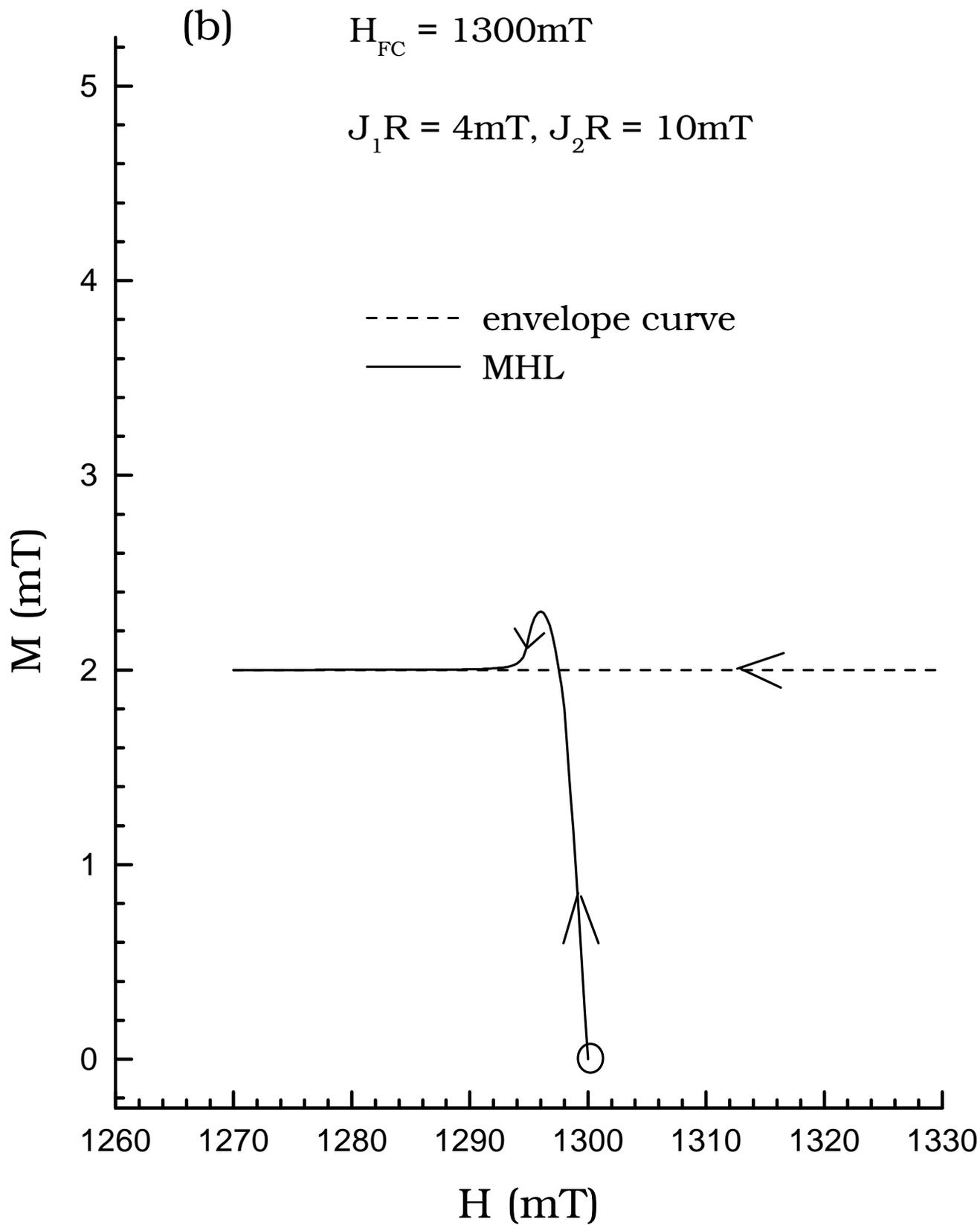
\end{document}